# SOC Testing Methodology and Practice

Cheng-Wen Wu

Department of Electrical Engineering

National Tsing Hua University

Hsinchu, Taiwan 30013, ROC

**Abstract**—*On a commercial digital still camera (DSC) controller chip we practice a novel SOC test integration platform, solving real problems in test scheduling, test IO reduction, timing of functional test, scan IO sharing, embedded memory built-in self-test (BIST), etc. The chip has been fabricated and tested successfully by our approach. Test results justify that short test integration cost, short test time, and small area overhead can be achieved. To support SOC testing, a memory BIST compiler and an SOC testing integration system have been developed.*

## 1 Introduction

It is generally agreed that, for an SOC, the design and test engineers usually have to test the cores with very limited knowledge of the core test information. The issues of core access and isolation are being addressed by the IEEE P1500 standard [1]. Although the standard provides unified core test access methods, the test controller, test architecture, test access mechanism (TAM), and test integration are left to the SOC integrator. In [2], we stress two major issues in practical SOC test integration: 1) session-based test scheduling considering not only the realistic test control architecture and TAM bus, but also test IO limit; and 2) the coexistence of scan test and functional test for logic cores. A commercial digital still camera (DSC) controller SOC has been developed using our test integration platform—SOC Test Aid Console (STEAC). Results from the DSC controller chip show that our approach is more effective than others that use non-session-based test scheduling. Also, both the scan and functional tests are supported. All the tens of embedded memories are tested by the built-in self-test (BIST) circuits generated by our memory BIST compiler, BRAINS [3].

## 2 SOC Test Aid Console (STEAC)

Figure 1 shows the SOC test integration system called STEAC—SOC Test Aid Console [4], which consists of four modules: the STIL Parser, Core Test Scheduler, Test Insertion Tool, and Pattern Translator. The STIL Parser parses the test information of each IP. The test information is written in STIL and is generated by commercial ATPG tools. Therefore, STEAC can be integrated into a typical design flow easily. The test information includes the IO ports, scan structure (number of scan chains, length of each scan chain, etc.), and test vectors. With these core test information, Core Test Scheduler will schedule the core tests to reduce the overall test time. The Scheduler partitions core tests into several test sessions, and assigns the TAM wires to each core to meet the power and IO resource constraints. If the IP is a soft core, the scan chains can be reconfigured. The Core Test Scheduler will then rebalance scan chains for each assigned TAM width. The results can be fed back to the SOC integrator to reconfigure the scan chains to balance the chain length. The scheduling results are also used to generate the Test Controller, TAM bus, and Test Wrapper. Finally, the generated test circuitry is inserted into the original SOC netlist automatically. A new SOC design with DFT will be ready in minutes. The core test patterns are generated at the core level. After the cores are wrapped, the test patterns must be translated to the wrapper level and then to the chip level. The test patterns are cycle based, which can be applied by external ATE easily.

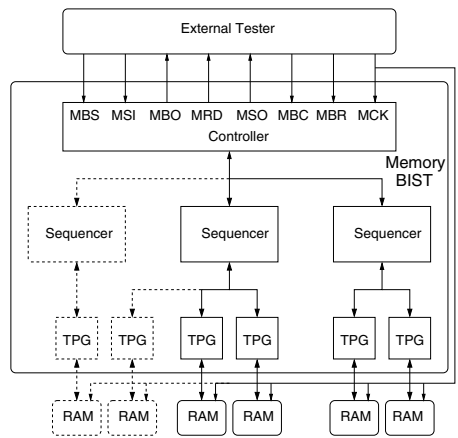

Figure 2: BIST architecture for multiple memory cores [5].

In Fig. 2 [5] we show our memory BIST architecture, which supports testing of plural heterogeneous memory cores. The tester can access all the on-chip memories via a single shared BIST Controller, while one or more Sequencers can be used to generate March-based test algorithms. Each Test Pattern Generator (TPG) attached to the memory will translate the March-based test commands to the respective RAM signals. With our automatic memory BIST generation system, BRAINS [3], one can generate the BIST circuit using the GUI or command shell, and evaluate the memory test efficiency among different designs easily. Moreover, BRAINS can be integrated with a memory compiler to deliver BISTed memory cores.

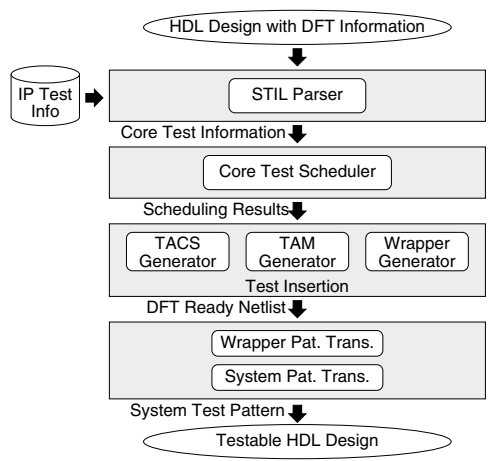

Figure 1: Test integration flow of STEAC [4].

## 3  Experimental Results

A DSC test chip has been implemented and fabricated to verify the proposed approach. This test chip is implemented with a standard 0.25$\mu$m CMOS technology. The major digital part of the chip includes a processor, JPEG codec, TV encoder, USB, external memory interface, and tens of single-port and two-port synchronous SRAMs with different sizes. Figure 3 [2] gives the block diagram of this test chip. The details are given in a companion paper [6].

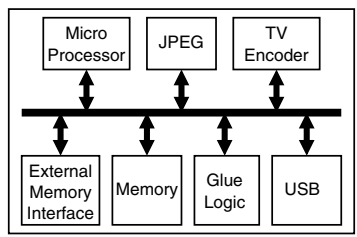

Figure 3: Block diagram of the test chip [2].

The IPs to be wrapped in this test chip include the USB, TV encoder, and JPEG cores. The USB core has 4 clock domains, 3 reset signals, 1 scan enable (SE) signal, and 6 test signals. There are 4 scan chains with dedicated scan input and output for each clock domain. The TV encoder has both scan and functional tests. The test pins include one clock, reset, SE, and test enable signals. There are two scan chains in the TV encoder, where one scan chain shares the output with a functional output. The legacy JPEG core has only functional patterns and one clock domain. The clock signals for the IPs are generated by an internal PLL. The detailed test information of the IPs are shown in Table 1 [2]. In the scan-test mode, the USB core has 2 scan vectors that are enabled by SE during Pulse-Clock, while the TV encoder has only one.

Table 1: Test information of the cores [2].

| Core | TI | TO | PI | PO | Scan chains (Lengths) | Patterns (Type) |
|---|---|---|---|---|---|---|
| USB | 18 | 4 | 221 | 104 | 4 (1,629, 78, 293, 45) | 716 (Scan) |
| TV | 6 | 1 | 25 | 40 | 2 (577, 576) | 229 (Scan) / 202,673 (Func.) |
| JPEG | 1 | 0 | 165 | 104 | No scan | 235,696 (Func.) |

When the test IO resource constraint is considered, parallel testing may not be better than serial testing. This is because more test control IOs are needed for parallel testing, so fewer IO pins can be used as the test data IOs (i.e., TAM IOs). Since there are also cases when parallel testing leads to shorter test time than serial testing, it is important to take chip IO pins into consideration so far as test time evaluation is concerned. In the DSC case, we tried several scheduling approaches, and found that the session-based approach (with three test sessions) has the shortest total test time— 4,371,194 clock cycles as opposed to 4,713,935 cycles by non-session-based approach. The total test IOs of the three large cores are 19, including 6 clock signals, 4 reset signals, 7 test enable signals, and 2 SE signals. With shared test IOs, the test control IO counts are reduced. With STEAC, the Test Wrappers, TAM, and Test Controller have been automatically generated and inserted into the original test chip design in 5 minutes, using a SUN Blade 1000 workstation with dual 750MHz processors and 2GB RAM. The area of the WBR cell is equivalent to 26 two-input NAND gates. The Test Controller and TAM multiplexer require about 371 and 132 gates, respectively—their hardware overhead is only about 0.3%. Again, all the embedded memories are tested with the BIST circuits generated by BRAINS, which has been integrated into STEAC, as shown in Fig. 4.

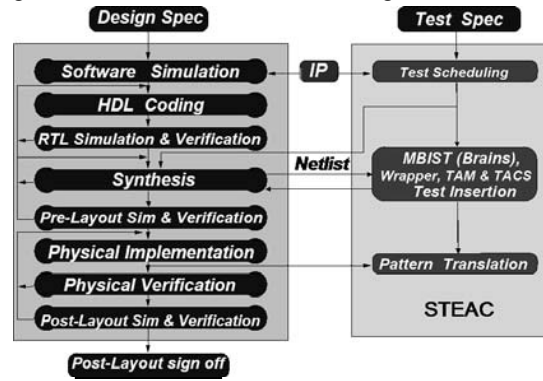

Figure 4: Integration of BRAINS into STEAC.

## 4  Conclusions

We have presented an SOC test integration platform, with complete solutions for real problems in test scheduling, test IO reduction, functional test, scan IO sharing, embedded memory BIST, etc. The fabricated test chip has been verified, and the test design has been successfully implemented on a commercial DSC chip. Test results justify that short test integration cost, short test time, and small area overhead can be achieved.

This work has been the result of the effort of many people, including my students who have coauthored the papers with me, as listed in the references.

## References

[1] E. Marinissen, R. Kapur, and Y. Zorian, "On using IEEE P1500 SECT for test plug-n-play", in *Proc. Int. Test Conf. (ITC)*, 2000, pp. 770–777.

[2] K.-L. Cheng, J.-R. Huang, C.-W. Wang, C.-Y. Lo, L.-M. Denq, C.-T. Huang, C.-W. Wu, S.-W. Hung, and J.-Y. Lee, "An SOC test integration platform and its industrial realization", in *Proc. Int. Test Conf. (ITC)*, Charlotte, Oct. 2004.

[3] C. Cheng, C.-T. Huang, J.-R. Huang, C.-W. Wu, C.-J. Wey, and M.-C. Tsai, "BRAINS: A BIST complier for embedded memories", in *Proc. IEEE Int. Symp. Defect and Fault Tolerance in VLSI Systems (DFT)*, Yamanashi, Oct. 2000, pp. 299–307.

[4] C.-W. Wang, J.-R. Huang, K.-L. Cheng, H.-S. Hsu, C.-T. Huang, C.-W. Wu, and Y.-L. Lin, "A test access control and test integration system for system-on-chip", in *Sixth IEEE Int. Workshop on Testing Embedded Core-Based System-Chips (TECS)*, Monterey, California, May 2002, pp. P2.1–P2.8.

[5] K.-L. Cheng, C.-M. Hsueh, J.-R. Huang, J.-C. Yeh, C.-T. Huang, and C.-W. Wu, "Automatic generation of memory built-in self-test cores for system-on-chip", in *Proc. Tenth IEEE Asian Test Symp. (ATS)*, Kyoto, Nov. 2001, pp. 91–96.

[6] C.-L. Chen, J.-Y. Lin, and Y.-L. Lin, "Integration, verification and layout of a complex multimedia SOC", in *Proc. Design, Automation and Test in Europe (DATE)*, Munich, Mar. 2005 (to appear).